\pdfoutput=1
\documentclass[conference]{IEEEtran}
\IEEEoverridecommandlockouts

\usepackage{cite}
\usepackage{amsmath,amssymb,amsfonts}
\usepackage{graphicx}
\usepackage{xcolor}
\usepackage{booktabs}
\usepackage{multirow}
\usepackage{siunitx}
\usepackage{subcaption}
\usepackage{url}
\usepackage{placeins}
\usepackage{dblfloatfix}
\usepackage{float}
\usepackage[hidelinks]{hyperref}
\usepackage{orcidlink}

\graphicspath{{Figures/}}

\title{RespGeomLib: A Reproducible Parametric Engine for Generating Analysis-Ready Human Airway Lumen Geometry}

\author{\scriptsize
\IEEEauthorblockN{Nichula Wasalathilaka\orcidlink{0009-0003-5662-009X}}
\IEEEauthorblockA{\textit{University of Peradeniya}\\
Peradeniya, Sri Lanka\\
e20425@eng.pdn.ac.lk}
\and
\IEEEauthorblockN{Parakrama Ekanayake\orcidlink{0000-0002-5639-8105}}
\IEEEauthorblockA{\textit{University of Peradeniya}\\
Peradeniya, Sri Lanka\\
mpbe@eng.pdn.ac.lk}
\and
\IEEEauthorblockN{Roshan Godaliyadda\orcidlink{0000-0002-3495-481X}}
\IEEEauthorblockA{\textit{University of Peradeniya}\\
Peradeniya, Sri Lanka\\
roshang@eng.pdn.ac.lk}
}

\begin{document}
\maketitle

% -------------------- Abstract & Keywords --------------------
\begin{abstract}
CT-derived airway models support pulmonary morphometry and airflow simulation, but are often limited by distal scan resolution and the need for substantial cleanup near bifurcations. Procedural alternatives are reproducible, yet many rely on stitched tubular primitives that introduce non-smooth junctions and poorly defined open boundaries. We present \textbf{RespGeomLib}, a reproducible parametric engine for generating analysis-ready human airway lumen surfaces from compact YAML specifications. The framework combines port-based assembly with implicit smooth-min junction blending to produce seamless junctions, while avoiding full-tree voxelization through analytic segments and local implicit extraction around bifurcations. Quantitatively, RespGeomLib yields cleaner junctions than a Boolean/stitch baseline and is substantially faster and more memory-efficient than whole-tree global implicit extraction. We further demonstrate morphometry-guided tree generation, controlled synthetic airway variants, and CFD-ready export with stable airflow simulation. RespGeomLib targets biomedical workflows requiring reproducible morphometry, controlled synthetic variants, and simulation-ready lumen geometry.The code is publicly available at \url{https://nichula01.github.io/Respgeomlib/}.

\end{abstract}

% -------------------- Main Sections --------------------
\begin{figure*}[t]
  \centering
  \includegraphics[width=0.95\textwidth]{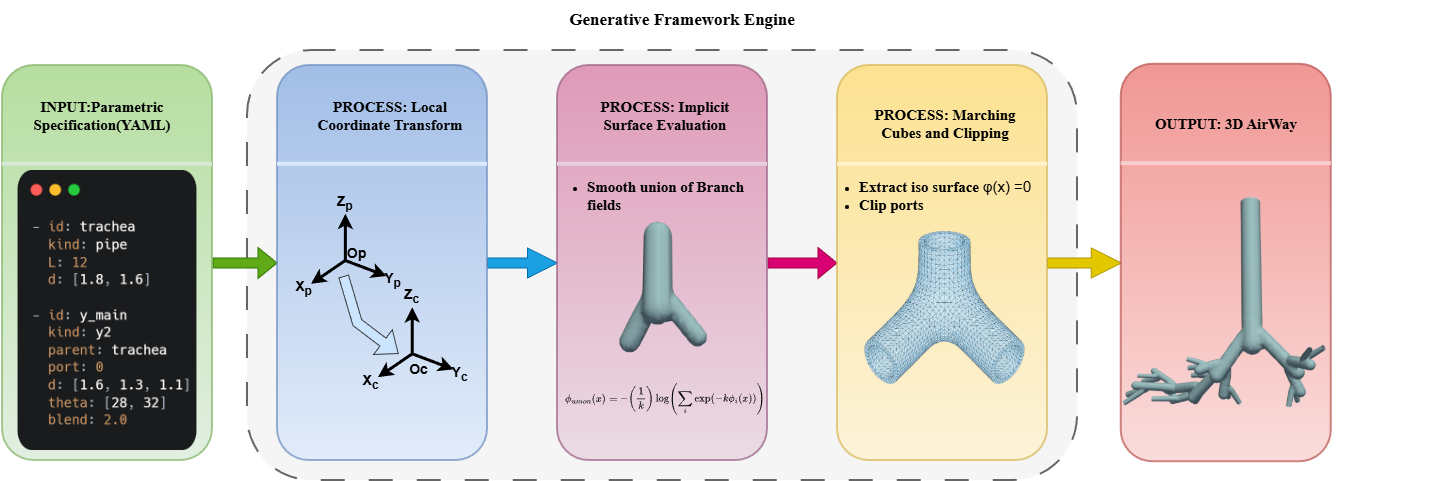}
  \caption{Overview of the RespGeomLib pipeline. A compact YAML specification defines primitives and parent--child attachments via ports. Segments are placed using local-to-world frame transforms; junctions are formed via a blended implicit field; the $\phi(\mathbf{x})=0$ surface is extracted locally using marching cubes, clipped at ports, and cleaned to produce an analysis-ready lumen mesh.}
  \label{fig:pipeline_overview}
\end{figure*}

\section{Introduction}
\label{sec:intro}

Quantitative analysis of the human airway tree underpins pulmonary health analytics, including the study of pathological remodeling in obstructive lung diseases and computational investigations of ventilation, aerosol transport, and airflow redistribution~\cite{Pal_gyi_2006,Pan_2025,James_2007,Bartlett_2023,Tang_2011}. In current practice, airway geometries are often obtained from CT via segmentation, centerline extraction, and surface reconstruction~\cite{Salama_2021,Yoo_2021}. Although such reconstructions are anatomically grounded, they remain sensitive to image noise and partial-volume effects, are fundamentally limited by scan resolution in distal generations~\cite{Kirby_2023}, and often require substantial cleanup near bifurcations before reliable morphometric analysis or CFD. Classic morphometric priors such as Weibel-style generation statistics and the ICRP respiratory tract model remain important as reproducible references for rule-based airway construction and extension~\cite{Lorenz_1966,Taylor_1996}.

Procedural modeling offers a reproducible alternative when distal branches are not directly resolved from imaging. Many existing approaches represent airways primarily as centerlines~\cite{Tawhai_2000} or generate geometry by sweeping or stitching tubular primitives. While computationally efficient, such stitched constructions can introduce sharp seams, non-smooth junctions, and poorly defined open boundaries that degrade downstream measurement and simulation quality~\cite{Espinosa_Moreno_2023}. Implicit surface modeling provides a complementary alternative by representing geometry through scalar fields and extracting smooth unions via level sets~\cite{Blinn_1982,Wyvill_1986,Lorensen_1987}. However, applying implicit sampling over an entire airway tree can become costly because of the large spatial extent and strong scale variation from proximal to distal branches.

In this work, we present \textbf{RespGeomLib}, a reproducible parametric geometry engine for generating analysis-ready airway lumen surfaces from compact, human-readable specifications. Rather than targeting only a single downstream task, RespGeomLib is designed as a reusable engine for airway geometry generation that supports CFD studies, controlled synthetic-variant generation, and morphometry-guided tree synthesis from the same modeling abstraction. The library combines three key ideas: (i) a modular port-based assembly interface for hierarchical tree construction, (ii) implicit smooth-min junction blending to avoid stitched seams and internal walls, and (iii) a workflow that preserves clean, open inlet/outlet boundary ports for downstream analytics and simulation.

A central design goal of RespGeomLib is to retain the geometric flexibility of implicit modeling without incurring the cost of full-tree voxelization. To this end, the engine adopts a hybrid extraction strategy: straight and tapered segments are generated analytically, while implicit isosurface extraction is applied only locally on tight grids around junctions. This local-implicit formulation yields smooth bifurcations and trifurcations while remaining substantially more tractable than evaluating a single implicit field over the full airway tree volume. In addition, the same framework supports controllable localized lumen perturbations, enabling reproducible stenotic and dilated variants from a fixed base geometry while preserving topology and port definitions.

\paragraph{Contributions.}
\begin{itemize}
    \item RespGeomLib: a YAML-based parametric airway geometry engine that generates triangulated lumen meshes with explicit inlet/outlet ports from compact, human-readable specifications.
    \item Smooth junction construction: a smooth-min implicit formulation for bifurcations and trifurcations that yields seamless, lumen-continuous junctions without stitched seams or internal walls.
    \item Efficient hybrid meshing: a local-implicit/analytic pipeline that avoids full-tree voxelization and is supported by quantitative junction-quality and local-vs-global scaling benchmarks.
    \item Reusable downstream utility:demonstrations of controlled synthetic variants, morphometry-guided tree generation, and CFD-ready export from the same geometry engine.
\end{itemize}

% \input{chapters/02_related_work}
% ============================================================
% methodology.tex  (MERCon final version)
% ============================================================

\section{Proposed Geometry Method}
\label{sec:method}

\subsection{Overview and design goals}
\label{sec:method_overview}
We present a reproducible airway-geometry pipeline that generates analysis- and simulation-ready lumen surfaces from compact parametric specifications and assembles them into a branching tree (Fig.~\ref{fig:pipeline_overview}). The design goals are threefold: (i) \emph{modularity}, through explicit inlet/outlet \emph{ports} that support hierarchical assembly and extension; (ii) \emph{smooth junction construction}, using implicit blending rather than sharp Boolean seams; and (iii) \emph{analysis/CFD-readiness}, through continuous lumen surfaces with well-defined open boundary ports for boundary-condition assignment. Beyond manually specified central airways, the same framework supports rule-based growth using published morphometry trends (e.g., Weibel/ICRP) with optional Murray-type diameter constraints.

\subsection{Geometry primitives and implicit junction generation}
\label{sec:geom_primitives}

\begin{figure*}[!t]
  \centering

  % ---------- Left: building-block figure ----------
  \begin{minipage}[t]{0.43\textwidth}
    \centering
    \vspace{0pt}
    \includegraphics[
      width=\linewidth,
      trim=15 10 15 10,
      clip
    ]{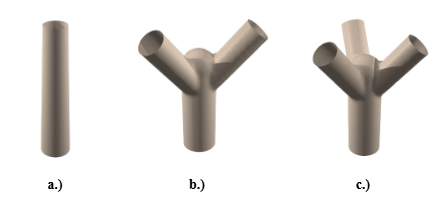}

    \vspace{1mm}
    {\small\textbf{(a) RespGeomLib building blocks}}
  \end{minipage}
  \hfill
  % ---------- Right: parametric input table ----------
  \begin{minipage}[t]{0.53\textwidth}
    \centering
    \vspace{0pt}
    \scriptsize
    \renewcommand{\arraystretch}{1.15}
    \setlength{\tabcolsep}{4pt}

    \resizebox{\linewidth}{!}{%
    \begin{tabular}{@{}lll@{}}
    \toprule
    \textbf{Block} & \textbf{Component} & \textbf{Parameters} \\
    \midrule

    Pipe
    & Tapered tube
    & Length $L$; diameters $(d_{\mathrm{in}}, d_{\mathrm{out}})$ \\

    \midrule

    Y2
    & \textit{Trunk}
    & Length $L_t$; diameter $d_t$ \\

    & \textit{Children} ($i=1,2$)
    & Length $L_i$; diameter $d_i$; angles $(\theta_i,\phi_i)$ \\

    \midrule

    Y3
    & \textit{Trunk}
    & Length $L_t$; diameter $d_t$ \\

    & \textit{Children} ($i=1,2,3$)
    & Length $L_i$; diameter $d_i$; angles $(\theta_i,\phi_i)$ \\

    \bottomrule
    \end{tabular}
    }

    \vspace{1mm}
    {\small\textbf{(b) Parametric inputs}}
  \end{minipage}

  \caption{RespGeomLib building blocks and their geometric inputs. 
  (a) Analytic pipe segment, two-way junction (Y2), and three-way junction (Y3). 
  (b) Parametric inputs used to define each primitive block.}
  \label{fig:building_blocks_primitives}
\end{figure*}

\begin{figure*}[t]
  \centering
  \includegraphics[width=0.95\textwidth]{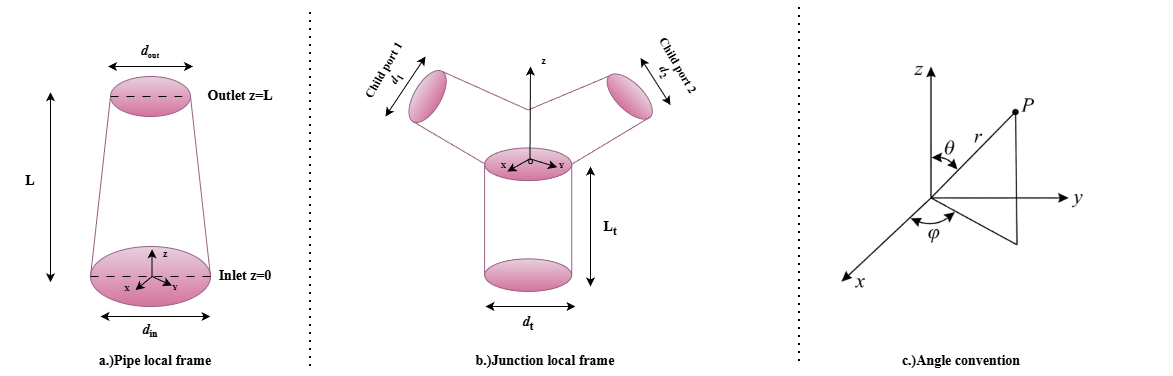}
  \caption{Local coordinate conventions. (a) Pipe: axis aligned with $+\hat{\mathbf{z}}$, inlet at $z=0$, outlet at $z=L$, optional taper $d_{\mathrm{in}}\!\rightarrow d_{\mathrm{out}}$. (b) Junction: centerline intersection at the origin, trunk along $-\hat{\mathbf{z}}$, and child ports defining attachment directions. (c) Angles: elevation $\theta$ (from $+\hat{\mathbf{z}}$) and azimuth $\phi$ in the $xy$-plane from $+\hat{\mathbf{x}}$.}
  \label{fig:local_conventions}
\end{figure*}

RespGeomLib supports three parametric primitives Pipe, Y2, and Y3 summarized and illustrated in Fig.~\ref{fig:building_blocks_primitives}. Each primitive is defined in a canonical local frame with circular cross-sections and explicit inlet/outlet ports, and is mapped into the global tree by attaching a child to a parent port (Sec.~\ref{sec:frames_tree}). A Pipe is aligned with local $+\hat{\mathbf{z}}$ and may taper from $d_{\mathrm{in}}$ to $d_{\mathrm{out}}$ over length $L$. For junctions, the centerline intersection is placed at the origin; the trunk follows $-\hat{\mathbf{z}}$; and child directions are specified by elevation--azimuth angles $(\theta_i,\phi_i)$ (Fig.~\ref{fig:local_conventions}).

\subsubsection{Implicit blended junction surface (local extraction)}
\label{sec:implicit_junction}
Junctions are modeled as the zero level-set of a blended implicit tube field. For branch $i$ with centerline $C_i$ and radius $r_i$, we define
\begin{equation}
\phi_i(\mathbf{x}) = d(\mathbf{x},C_i) - r_i,
\label{eq:tube_field}
\end{equation}
where $d(\mathbf{x},C_i)=\min_{\mathbf{y}\in C_i}\|\mathbf{x}-\mathbf{y}\|_2$ is the Euclidean distance from point $\mathbf{x}$ to centerline $C_i$. The soft-union field is then
\begin{equation}
\phi(\mathbf{x})=
-\frac{1}{\kappa}\log\!\left(\sum_{i=1}^{N} e^{-\kappa \phi_i(\mathbf{x})}\right),
\qquad
\mathcal{S}=\{\mathbf{x}:\phi(\mathbf{x})=0\},
\label{eq:softmin_union}
\end{equation}
where $\kappa>0$ controls blend sharpness: larger $\kappa$ yields a sharper union, while smaller $\kappa$ produces smoother transitions between branches. The surface $\mathcal{S}$ is extracted by sampling $\phi$ on a tight \emph{local} grid around each junction and applying marching cubes. Straight and tapered segments remain analytic, thereby avoiding full-tree voxelization.

\subsection{Coordinate frames and tree assembly}
\label{sec:frames_tree}
Assembly attaches each child primitive to a parent port represented by a \emph{port frame}, whose origin is the port center and whose $\hat{\mathbf{z}}$ axis is aligned with the outward centerline direction. Child directions are specified by elevation--azimuth angles $(\theta,\phi)$ in the parent port frame and mapped to the unit vector
\begin{equation}
\hat{\mathbf d}(\theta,\phi)=
\begin{bmatrix}
\sin\theta\cos\phi\\
\sin\theta\sin\phi\\
\cos\theta
\end{bmatrix}.
\label{eq:dir_angles}
\end{equation}
The desired child axis is then set as $\hat{\mathbf z}_c=\hat{\mathbf d}(\theta,\phi)$ expressed in the parent frame.

\subsubsection{Minimal-twist child frame}
\label{sec:minimal_twist}
Given parent axes $(\hat{\mathbf x}_p,\hat{\mathbf y}_p,\hat{\mathbf z}_p)$ and desired child axis $\hat{\mathbf z}_c$, we construct a minimal-twist child frame by projecting $\hat{\mathbf x}_p$ onto the plane orthogonal to $\hat{\mathbf z}_c$:
\begin{equation}
\tilde{\mathbf x}=\hat{\mathbf x}_p-(\hat{\mathbf x}_p^\top\hat{\mathbf z}_c)\hat{\mathbf z}_c,
\qquad
\hat{\mathbf x}_c=\frac{\tilde{\mathbf x}}{\|\tilde{\mathbf x}\|}.
\label{eq:min_twist}
\end{equation}
If $\|\tilde{\mathbf x}\|$ is near zero, we instead project $\hat{\mathbf y}_p$ using the same rule. The basis is completed by
\begin{equation}
\hat{\mathbf y}_c=\hat{\mathbf z}_c\times\hat{\mathbf x}_c,
\end{equation}
which yields a consistent local frame with minimal spin about the branch axis.

\subsubsection{Local-to-world placement}
\label{sec:local_to_world}
Each primitive is generated in local coordinates and mapped to world coordinates by a rigid transform with rotation matrix
$R=[\hat{\mathbf x}_c\ \hat{\mathbf y}_c\ \hat{\mathbf z}_c]$ and origin $\mathbf{o}$:
\begin{equation}
\mathbf{x}_{\text{world}}=R\,\mathbf{x}_{\text{local}}+\mathbf{o}.
\label{eq:rigid_transform}
\end{equation}
The transformed segment meshes are then merged to form the full lumen surface.

\subsection{Controlled synthetic variants via parametric primitives}
\label{sec:pathology_primitives}
To support reproducible synthetic-variant generation, we model localized lumen remodeling along a chosen normalized centerline coordinate $s\in[0,1]$. Given baseline radius $r_0(s)$, the perturbed radius is
\begin{equation}
r(s)=r_0(s)\bigl(1+\alpha\,w(s)\bigr),
\label{eq:pathology_radius_mod}
\end{equation}
where $w(s)$ is a smooth localized window (e.g., Gaussian) and $\alpha$ controls perturbation severity. Negative $\alpha$ yields \emph{stenosis} (localized narrowing) and positive $\alpha$ yields \emph{dilation} (localized widening). Because the perturbation is applied to a fixed base geometry while preserving topology and port placement, the same mechanism enables labeled synthetic geometry cohorts for controlled benchmarking and downstream CFD studies.

\subsection{Implementation details and export}
\label{sec:impl_export}

\subsubsection{Local grid sizing for implicit extraction}
\label{sec:grid_sizing}
Implicit extraction is performed \emph{per junction} on a uniform grid over a tight local bounding box around the trunk and child centerlines, expanded by margin $m=2r_{\max}$ to avoid truncation of the $\phi=0$ surface. Grid spacing is defined relative to local radii:
\begin{equation}
h = \frac{r_{\max}}{\eta},
\label{eq:grid_spacing}
\end{equation}
where $r_{\max}$ is the maximum branch radius in the junction and $\eta$ is a user-defined sampling density (samples per radius). This local-grid strategy is the key mechanism used to avoid the cost of evaluating a single implicit field over the full airway tree volume.

\subsubsection{Open-port enforcement and validity checks}
\label{sec:open_ports}
The lumen is intentionally \emph{open} at prescribed inlet/outlet ports for boundary-condition assignment. After assembly, we clip the surface at port planes to enforce clean planar boundary loops, apply deterministic cleanup (duplicate/degenerate removal and consistent normals), and validate manifoldness. The only boundary edges should coincide with intended port loops; any unintended cracks or holes are treated as validity failures.

\subsubsection{Biologically constrained growth rules}
\label{sec:bio_rules}
Beyond manually specified proximal airways, RespGeomLib supports automatic extension using published morphometry trends, including generation-wise diameter and length patterns from Weibel/ICRP models, with optional Murray-type branching constraints. This produces plausible subtrees while preserving the same port-based interface and deterministic build logic.

\subsubsection{Tree specification from YAML}
\label{sec:yaml_tree}
Trees are specified in a human-readable YAML format in which each segment declares its primitive type (Pipe/Y2/Y3), geometric parameters, and parent reference via $(\texttt{parent\_id},\texttt{parent\_port\_index})$. The loader resolves a deterministic build order, and the builder attaches each segment using Eqs.~\eqref{eq:dir_angles}--\eqref{eq:rigid_transform}. This compact specification makes the engine reproducible, editable, and suitable for scripted generation of airway families and controlled variants.
% ============================================================
% results.tex  (MERCon final version, quantitative text added)
% Keeps current figure/table placement strategy unchanged
% ============================================================

\section{Results}
\label{sec:results}

\subsection{Junction-quality benchmark}
\label{sec:results_junction_quality}

To quantify the geometric advantage of smooth local implicit construction, we compared a matched Y2 junction generated by a Boolean/stitch baseline against the \textsc{RespGeomLib} smooth-min formulation. Table~\ref{tab:junction_quality} shows that both methods preserved the intended three open ports, but the stitched baseline exhibited 6 non-manifold edges, large port non-planarity (maximum error 5.9796), and failed a best-effort planar capping test. In contrast, \textsc{RespGeomLib} produced 0 non-manifold edges, near-planar ports (maximum error 0.0010), and successfully capped to a watertight surface. These results support the use of smooth local implicit blending for generating cleaner analysis- and CFD-ready junctions.

\begin{table}[t]
\centering
\small
\setlength{\tabcolsep}{4pt}
\caption{Junction-quality benchmark for a Boolean/stitch baseline versus the RespGeomLib smooth-min Y2 junction.}
\label{tab:junction_quality}
\begin{tabular}{lrrrrr}
\toprule
Method & Loops & Non-man. & Planarity max & Cap OK & Cells \\
\midrule
Boolean/stitch & 3 & 6 & 5.9796 & no  & 7213  \\
RespGeomLib    & 3 & 0 & 0.0010 & yes & 63436 \\
\bottomrule
\end{tabular}
\end{table}
% If your file still uses the older typo, replace with:
% \input{tables/juntion_quality}

\subsection{Computational scaling of local versus global implicit extraction}
\label{sec:results_scaling}

We next evaluated the computational benefit of junction-local implicit extraction relative to a whole-tree global implicit baseline on balanced Y2 airway trees. Table~\ref{tab:scaling-benchmark} reports matched-depth comparisons. At depths 1--3, global extraction was 76.7$\times$, 201.8$\times$, and 279.8$\times$ slower than the local strategy, while also using 1.11$\times$--1.31$\times$ higher peak memory. The local method remained practical at larger tree depths, whereas the global baseline became expensive already at shallow depths. This confirms that restricting implicit sampling to tight local junction grids is the key tractability advantage of the hybrid meshing strategy.

\begin{table}[t]
\centering
\scriptsize
\setlength{\tabcolsep}{3pt}
\caption{Matched-depth scaling summary for balanced Y2 airway trees.}
\label{tab:scaling-benchmark}
\begin{tabular}{rrrrrrrr}
\toprule
Depth & Br. & L-RT (s) & G-RT (s) & RT$\times$ & L-Mem (MB) & G-Mem (MB) & Mem$\times$ \\
\midrule
1 & 3  & 0.074 & 5.640   & 76.7  & 135.4 & 150.5 & 1.11 \\
2 & 7  & 0.185 & 37.358  & 201.8 & 136.7 & 165.1 & 1.21 \\
3 & 15 & 0.420 & 117.519 & 279.8 & 140.6 & 183.9 & 1.31 \\
\bottomrule
\end{tabular}
\end{table}

\subsection{CT-derived versus procedural airway}
\label{sec:results_ct_qualitative}

\begin{figure}[!t]
    \centering
    \includegraphics[width=0.92\linewidth]{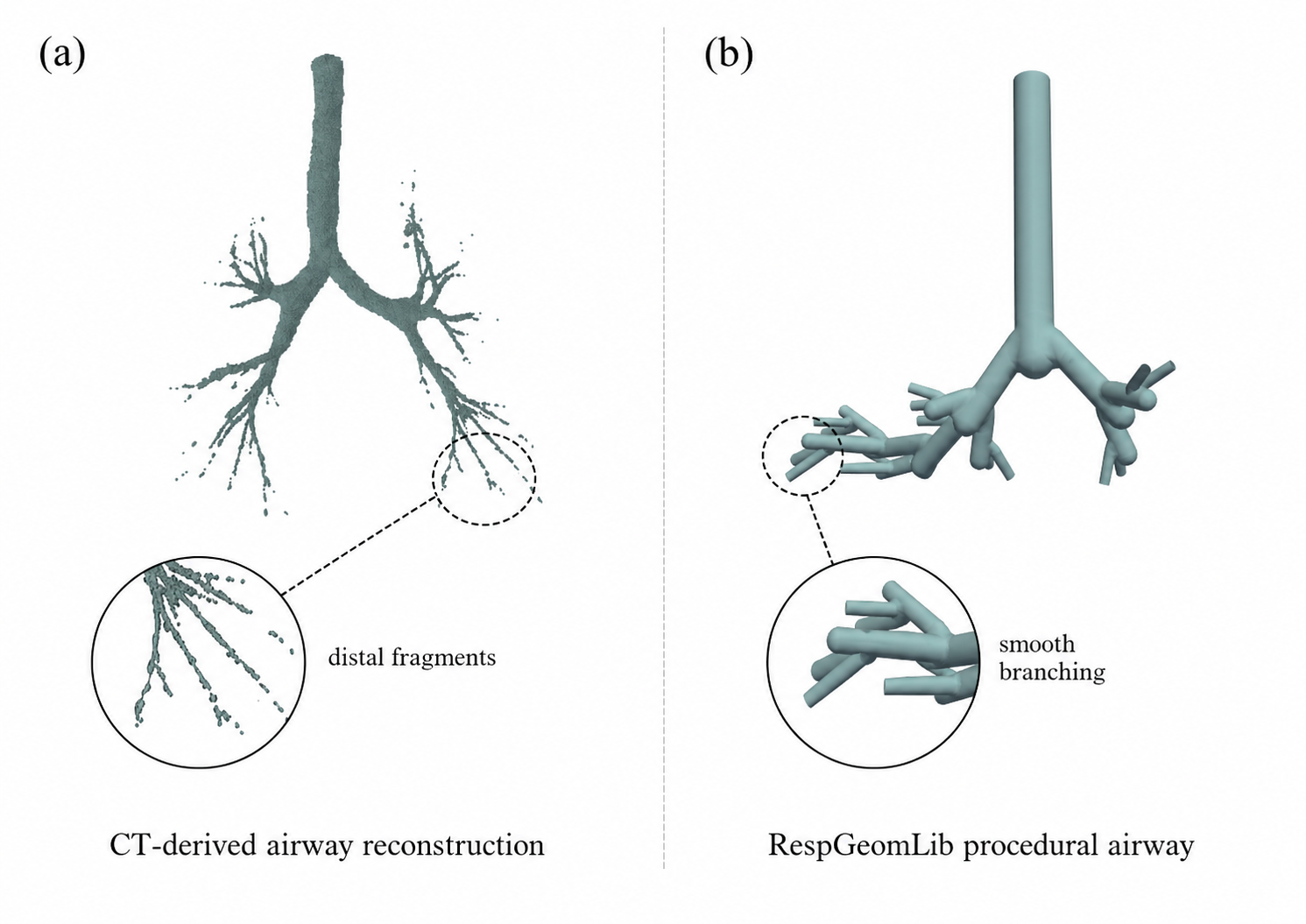}
    \caption{Qualitative comparison between (a) a CT-derived airway reconstruction and (b) a \textsc{RespGeomLib} procedural airway. The CT-derived surface shows distal fragments and irregular peripheral structure, whereas the procedural model provides smooth, controllable branching from an explicit parametric specification.}
    \label{fig:ct_vs_procedural}
\end{figure}

Fig.~\ref{fig:ct_vs_procedural} contrasts a CT-derived airway reconstruction with a \textsc{RespGeomLib} procedural airway of comparable proximal branching structure. The CT-derived surface exhibits distal fragments and irregular peripheral structure, whereas the procedural airway yields a smooth, controllable geometry with explicit branch and junction definitions. We include this comparison as a qualitative illustration of the trade-off between anatomically grounded reconstruction and reproducible parametric generation; the main quantitative evidence in this paper is provided by the junction-quality and scaling benchmarks.

% ============================================================
% IMPORTANT:
% Keep this early so LaTeX can place it at the top of the next page.
% Counter adjustment preserved exactly as in your current draft.
% ============================================================

\addtocounter{figure}{1}
\begin{figure*}[!t]
    \centering
    \includegraphics[width=0.92\textwidth]{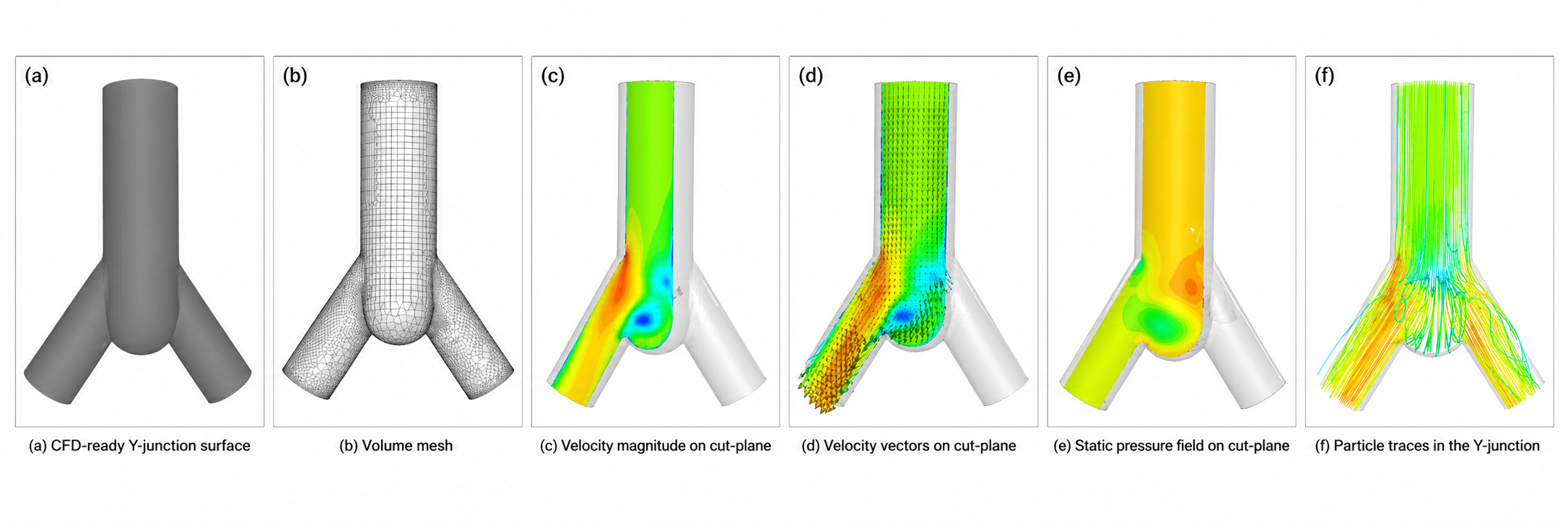}
    \caption{CFD demonstration on a \textsc{RespGeomLib} Y-junction in SimScale. (a) Exported CFD-ready closed surface. (b) Generated volume mesh. (c) Velocity magnitude on a representative cut-plane. (d) Velocity vectors on the same cut-plane. (e) Static pressure field on the same cut-plane. (f) Particle traces colored by velocity magnitude.}
    \label{fig:cfd_grid}
\end{figure*}
\addtocounter{figure}{-2}

\subsection{Controlled synthetic variants}
\label{sec:results_synthetic_variants}

\begin{figure}[H]
    \centering
    \includegraphics[width=0.82\linewidth]{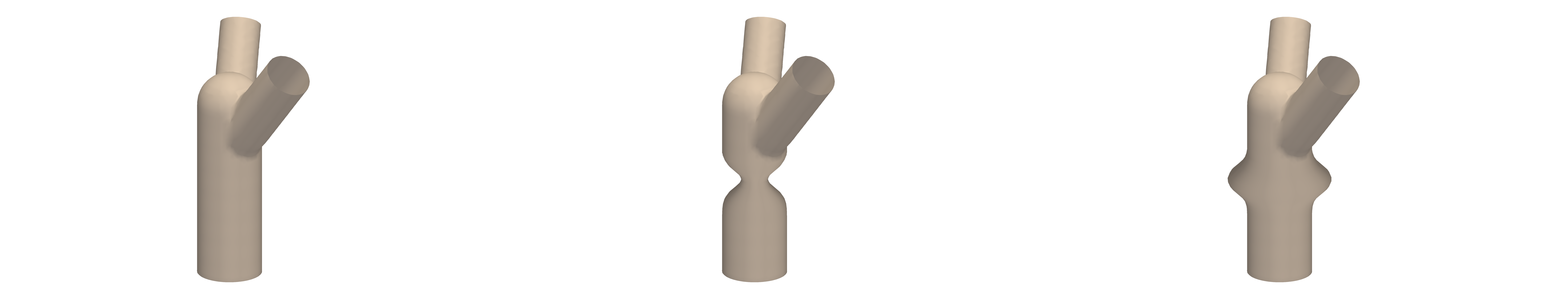}
    \caption{Implicit Y-junction with parametric stenosis/dilation ($\alpha \in \{0,-0.6,+0.6\}$) using Eq.~\eqref{eq:pathology_radius_mod}.}
    \label{fig:pathology_triptych}
\end{figure}

To demonstrate controlled synthetic-variant generation, we applied the localized radius-modulation model in Eq.~\eqref{eq:pathology_radius_mod} to a fixed implicit Y-junction template. Fig.~\ref{fig:pathology_triptych} shows baseline, stenotic, and dilated variants generated under identical settings. Because the perturbation is applied to a fixed base geometry while preserving topology and port definitions, the same mechanism supports reproducible severity-controlled synthetic cohorts for benchmarking, sensitivity analysis, and downstream CFD studies.

\subsection{CFD usability on a \textsc{RespGeomLib} Y-junction}
\label{sec:results_cfd}

To demonstrate downstream usability for airflow simulation, we exported a closed, CFD-ready Y-junction surface from \textsc{RespGeomLib} and solved a steady incompressible case in SimScale with one velocity inlet and two pressure outlets. The setup used steady incompressible RANS ($k$--$\omega$ SST), air $(\rho,\mu)=(1.225~\mathrm{kg\,m^{-3}},1.8\times10^{-5}~\mathrm{Pa\,s})$, inlet velocity $U_{\mathrm{in}}=0.2~\mathrm{m\,s^{-1}}$, outlet gauge pressure $p_g=0~\mathrm{Pa}$, and no-slip walls. A hex-dominant internal mesh with approximately $10^6$ cells was generated from the exported watertight surface. Fig.~\ref{fig:cfd_grid} shows the exported geometry, volume mesh, representative cut-plane fields, and particle traces, while Fig.~\ref{fig:cfd_residuals} summarizes normalized residual convergence over the run. Quantitatively, the case yielded an estimated Reynolds number consistent with the low-speed laminar-to-transitional airway regime, a mean inlet pressure of approximately $1.0\times10^{-1}$~Pa, a corresponding junction pressure drop of approximately $1.0\times10^{-1}$~Pa, and a slightly asymmetric outlet flow split of 46.5\% versus 53.5\%. The maximum velocity magnitude reached approximately $0.26~\mathrm{m\,s^{-1}}$, and the final residual range was on the order of $10^{-5}$ (Table~\ref{tab:cfd_metrics}). Overall, these results verify that exported junction geometries support standard meshing, boundary-condition assignment, and post-processing in a commercial CFD workflow while producing stable solver behavior and physically plausible flow redistribution.

% Reset counter so this becomes Fig. 7 because CFD figure above is Fig. 6.
\setcounter{figure}{6}
\begin{figure}[!t]
    \centering
    \includegraphics[width=0.92\linewidth]{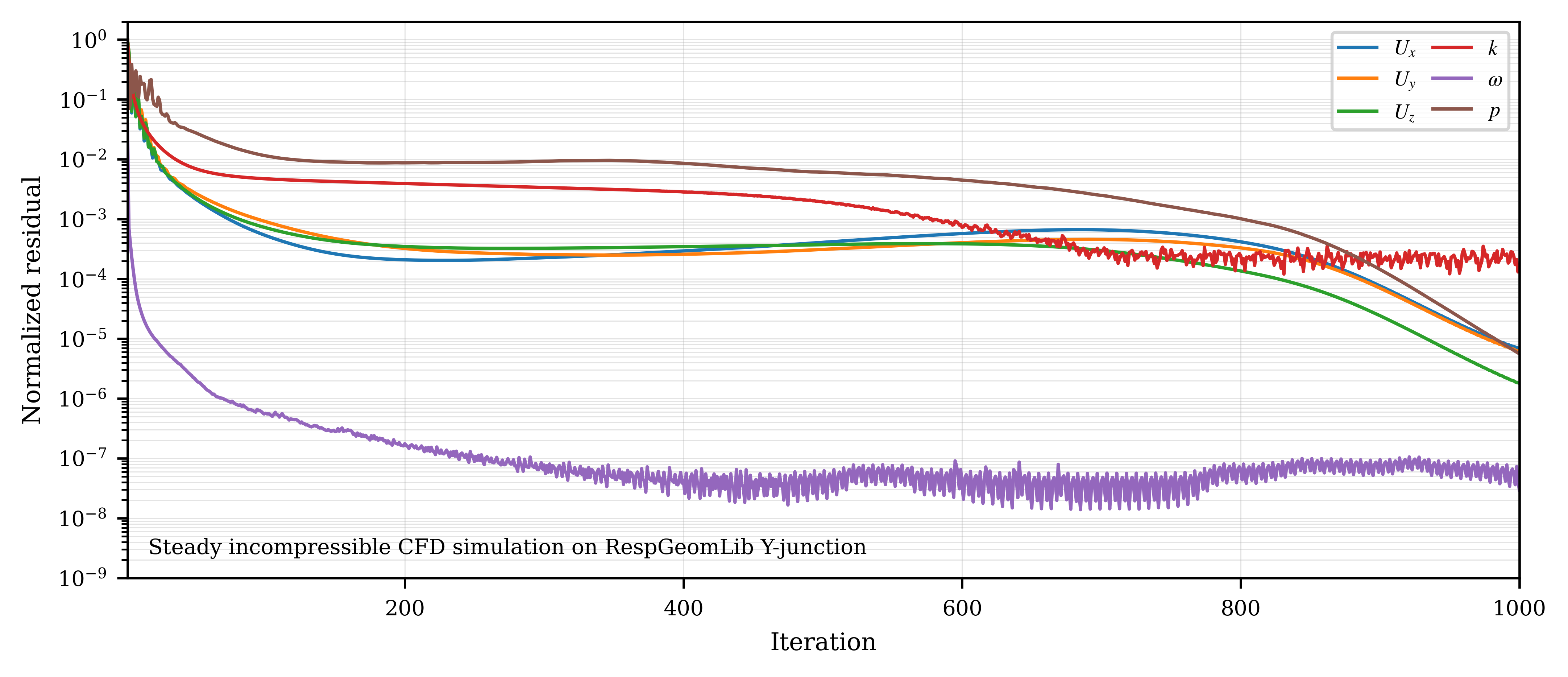}
    \caption{Normalized residual histories for the Y-junction CFD case, showing stabilization of the primary flow variables during the steady incompressible simulation.}
    \label{fig:cfd_residuals}
\end{figure}

\begin{table}[!t]
\caption{Quantitative CFD summary for the \textsc{RespGeomLib} Y-junction case.}
\label{tab:cfd_metrics}
\centering
\scriptsize
\setlength{\tabcolsep}{4pt}
\renewcommand{\arraystretch}{1.08}
\begin{tabular}{lc}
\toprule
Metric & Value \\
\midrule
Mesh cells & $\sim 10^{6}$ \\
Fluid density, $\rho$ & $1.225~\mathrm{kg\,m^{-3}}$ \\
Dynamic viscosity, $\mu$ & $1.8\times10^{-5}~\mathrm{Pa\,s}$ \\
Inlet velocity, $U_{\mathrm{in}}$ & $0.2~\mathrm{m\,s^{-1}}$ \\
Outlet gauge pressure, $p_g$ & $0~\mathrm{Pa}$ \\
Mean inlet pressure, $\bar{p}_{\mathrm{in}}$ & $1.0\times10^{-1}~\mathrm{Pa}$ \\
Mean outlet pressure, $\bar{p}_{\mathrm{out}}$ & $\approx 0~\mathrm{Pa}$ \\
Pressure drop, $\Delta p$ & $1.0\times10^{-1}~\mathrm{Pa}$ \\
Outlet-1 flow fraction & $46.5\%$ \\
Outlet-2 flow fraction & $53.5\%$ \\
Maximum velocity magnitude & $0.26~\mathrm{m\,s^{-1}}$ \\
Final residual range & $\sim 10^{-5}$ \\
\bottomrule
\end{tabular}
\end{table}

% Keep all remaining Results floats before Discussion.
\FloatBarrier
\section{Discussion}
\label{sec:discussion}

RespGeomLib is best understood as a reproducible \emph{geometry engine} for airway lumen generation rather than as a patient-specific reconstruction method. Its main value lies in providing a compact, editable, and simulation-ready representation that can be used consistently across multiple downstream workflows. The results support this positioning in two ways. First, the junction-quality benchmark shows that smooth local implicit blending yields cleaner open-port geometry than a stitched/Boolean baseline, with lower non-manifoldness, substantially improved port planarity, and successful capping to a watertight surface. Second, the scaling benchmark shows that restricting implicit extraction to tight local grids around junctions is substantially more tractable than evaluating a global implicit field over the full airway tree. Together, these findings support the core design choice of combining analytic segments with local implicit junction extraction.

The synthetic-variant mechanism further extends the utility of the engine by enabling controlled, labeled perturbations of a fixed base geometry. In this setting, stenotic and dilated variants can be generated while preserving topology and port definitions, making the framework suitable for reproducible sensitivity studies, benchmarking, and downstream CFD experiments under controlled geometric change. The CFD demonstration complements this by showing that exported \textsc{RespGeomLib} junctions can be meshed and simulated in a standard commercial workflow, yielding stable residual behavior and physically plausible flow fields.

\paragraph{Limitations.}
The current study focuses on idealized lumen geometry and a limited family of junction and tree configurations. Although the framework supports morphometry-guided growth and controlled perturbations, it does not yet model several factors that influence physiological realism, including patient-specific asymmetry, non-circular cross-sections, airway wall thickness, and fine-scale surface roughness. In addition, the CFD study is intended as a downstream usability demonstration rather than a comprehensive fluid-dynamics benchmark across multiple airway depths, flow regimes, or solver settings. Finally, the morphometry result reported here is best interpreted as a consistency check of the profile-driven generation pipeline rather than as an independent validation benchmark.

\section{Conclusion}
\label{sec:conclusion}

We presented \textsc{RespGeomLib}, a reproducible parametric engine for generating analysis-ready human airway lumen geometry from compact YAML specifications. The framework combines analytic pipe generation with local implicit smooth-blended junction construction, producing lumen meshes with explicit ports for downstream simulation and analysis. Compared with a stitched/Boolean junction baseline, \textsc{RespGeomLib} produces cleaner and more planar open-port geometry, and compared with whole-tree global implicit extraction, its local extraction strategy scales substantially better in runtime and memory.

Beyond baseline tree synthesis, the same engine supports morphometry-guided airway generation, controlled synthetic variants through localized stenosis/dilation primitives, and CFD-ready export for airflow studies. These capabilities make \textsc{RespGeomLib} a practical tool for reproducible geometry generation in health analytics workflows, including controlled benchmarking, synthetic dataset creation, and simulation studies. Future work will extend the framework toward richer anatomical variability, broader CFD validation, and larger airway-network experiments.

% -------------------- References --------------------
\bibliographystyle{IEEEtran}
\bibliography{references}

\end{document}